\documentclass[prl,twocolumn,showpacs,superscriptaddress,amsmath,amssymb]{revtex4-1}
\usepackage{graphicx}
\usepackage{subfigure}

\begin{document}

\title{{Persistent detwinning of iron pnictides by small magnetic fields}}
\author{S. Zapf} 
\altaffiliation{e-mail: sina.zapf@pi1.physik.uni-stuttgart.de}
\affiliation{1.~Physikalisches Institut, Universit\"at Stuttgart, Pfaffenwaldring 57, 70550 Stuttgart, Germany}
\author{C. Stingl}
\affiliation{I.~Physikalisches Institut, Universit\"at G\"ottingen, Friedrich-Hund-Platz 1, 37077 G\"ottingen, Germany}
\author{K. W. Post}
\affiliation{Department of Physics, University of California, San Diego, La Jolla, California 92093, USA}
\author{J. Maiwald}
\affiliation{I.~Physikalisches Institut, Universit\"at G\"ottingen, Friedrich-Hund-Platz 1, 37077 G\"ottingen, Germany}
\affiliation{Experimentalphysik~VI,  Universit\"at Augsburg, Universit\"atsstra{\ss}e 1, 86135 Augsburg, Germany}
\author{N. Bach}
\affiliation{I.~Physikalisches Institut, Universit\"at G\"ottingen, Friedrich-Hund-Platz 1, 37077 G\"ottingen, Germany}
\author{I. Pietsch}
\affiliation{I.~Physikalisches Institut, Universit\"at G\"ottingen, Friedrich-Hund-Platz 1, 37077 G\"ottingen, Germany}
\author{D. Neubauer}
\affiliation{1.~Physikalisches Institut, Universit\"at Stuttgart, Pfaffenwaldring 57, 70550 Stuttgart, Germany}
\author{A. L\"ohle}
\affiliation{1.~Physikalisches Institut, Universit\"at Stuttgart, Pfaffenwaldring 57, 70550 Stuttgart, Germany}
\author{C. Clauss}
\affiliation{1.~Physikalisches Institut, Universit\"at Stuttgart, Pfaffenwaldring 57, 70550 Stuttgart, Germany}
\author{S. Jiang}
\affiliation{1.~Physikalisches Institut, Universit\"at Stuttgart, Pfaffenwaldring 57, 70550 Stuttgart, Germany}
\author{H. S. Jeevan}
\affiliation{I.~Physikalisches Institut, Universit\"at G\"ottingen, Friedrich-Hund-Platz 1, 37077 G\"ottingen, Germany}
\affiliation{Department of Physics, PESITM, Sagar Road, 577204 Shimoga, India}
\author{D. N. Basov}
\affiliation{Department of Physics, University of California, San Diego, La Jolla, California 92093, USA}
\author{P. Gegenwart}
\affiliation{I.~Physikalisches Institut, Universit\"at G\"ottingen, Friedrich-Hund-Platz 1, 37077 G\"ottingen, Germany}
\affiliation{Experimentalphysik~VI,  Universit\"at Augsburg, Universit\"atsstra{\ss}e 1, 86135 Augsburg, Germany}
\author{M. Dressel}
\affiliation{1.~Physikalisches Institut, Universit\"at Stuttgart, Pfaffenwaldring 57, 70550 Stuttgart, Germany}

\date{\today}

\begin{abstract}
Our comprehensive study on EuFe$_2$As$_2$ reveals a dramatic reduction of magnetic detwinning fields compared to other $A$Fe$_2$As$_2$ ($A$ = Ba, Sr, Ca) iron pnictides by indirect magneto-elastic coupling of the Eu$^{2+}$ ions.
We find that only $\sim 0.1$\,T are sufficient for \emph{persistent} detwinning below the local Eu$^{2+}$ ordering; above $T_\text{Eu} = 19$\,K, higher fields are necessary.
Even after the field is switched off, a significant imbalance of twin domains remains constant up to the structural and electronic phase transition ($190$\,K).
This \emph{persistent} detwinning provides the unique possibility to study the low temperature electronic in-plane anisotropy of iron pnictides
without applying any symmetry-breaking external force.
\end{abstract}
\maketitle

The observation of a large in-plane anisotropy in iron pnictides has triggered tremendous research activity, as another potential key ingredient for high-temperature superconductivity was identified~\cite{Fisher11, Tanatar10, Yi11, Kasahara12, Chu2012, Dhital12, Jiang13, Rosenthal14, Fernandes14}. Similar to cuprates, the magnitude of the electronic anisotropy is unexpectedly large, because it notably surpasses the orthorhombic lattice distortion. In other words, the itinerant electrons do not just follow the lattice anisotropy, instead there is growing evidence for an underlying electronic ``nematic'' phase transition which breaks the crystal's rotational symmetry. 
As the formation of twin domains usually obscures this intrinsic anisotropy, sophisticated methods were already developed to detwin cuprates~\cite{Lavrov01, Lavrov02}. In the case of iron pnictides, the effect of typical laboratory magnetic fields on the Fe spins is rather weak~\cite{Chu10}. Thus, mechanical clamps are commonly used for detwinning single crystals~\cite{Fisher11}. However, this introduces an explicit symmetry breaking by uniaxial pressure, which must be considered carefully~\cite{Chu2012}: similar to ferromagnets, where the magnetization depends on the external magnetic field, the intrinsic nematic response can only be measured in the limit of zero  symmetry-breaking external force. Indeed, mechanical clamps were found to significantly enhance the transition temperatures in iron pnictides and even induce additional anisotropy above~\cite{Dhital12}. While those observations were easily accessible by comparing the response of twinned and mechanically detwinned crystals, at this point the influence of external forces on the low temperature anisotropy of iron pnictides is not clear at all. 

EuFe$_2$As$_2$ is a peculiar member of the 122 iron pnictides: similar to related high-temperature superconductors such as Ce 1111 pnictides and rutheno-cuprates~\cite{Chen08,Bauernfeind95}, the Eu$^{2+}$ spins order at low temperatures magnetically. Below $T_\text{Eu} = 19\,$K, they are arranged in an A-type antiferromagnetic structure, $i.e.$ within one layer ferromagnetically along the $a$-axis, but antiferromagnetically between neighboring layers (see Fig.~\ref{Fig4}, Ref.~\onlinecite{Xiao09,Jiang09}).
Furthermore, EuFe$_2$As$_2$ exhibits a structural and spin density wave (SDW) transition at $T_\text{s,SDW} = 190\,$K which can be suppressed by doping or pressure until superconductivity sets in at around $30\,$K. It is still under debate how superconductivity coexists with local Eu$^{2+}$ magnetism~\cite{Canfield11, Jeevan11, Cao11, Zapf13, Jiao11, Matsu11}.

Here we demonstrate the \emph{persistent} detwinning of EuFe$_2$As$_2$ by small magnetic fields, which yields similar detwinning fractions as commonly used mechanical devices. The detwinning mechanism is elucidated in detail by performing systematic resistivity, thermal expansion, magnetostriction, magneto-resistance, magneto-optical and magnetization measurements. Whereas previous studies~\cite{Xiao10} only reported that in-plane magnetic fields of the order 1\,T detwin EuFe$_2$As$_2$ at low temperatures with the crystal's longer $a$-axis parallel to the external field $H \parallel [110]_{\text{T}}$~\cite{tetra} and no persistent detwinning was revealed, we have investigated the magnetic detwinning in a broad temperature range, always following a well-defined cooling procedure: first, the sample was cooled from $T > T_\text{s,SDW}$ to low temperatures in zero magnetic field and the ``zero-field cooled'' (ZFC) response was measured. Afterwards, when an in-plane magnetic field parallel to the $[110]_{\text{T}}$-direction was first applied and then removed, we call this ``field treatment''~(FT). Reference measurements along the $[100]_{\text{T}}$-direction as well as details about the measurement techniques can be found in the Supplementary Information~\cite{Supplementary}.  We show that the magnetic detwinning with $a \parallel H$ can be achieved also above the Eu$^{2+}$ magnetic ordering temperature and that below $T_\text{Eu}$, an additional detwinning process with $b \parallel H$ takes place around 0.1\,T. Most strikingly, the detwinned state remains even when the field is switched off (below or above $T_\text{Eu}$) and the temperature is raised up to $T_\text{s,SDW}$. We suggest that the behavior is the result of an indirect coupling of the Eu$^{2+}$ spins to the lattice, which is mediated via the Fe$^{2+}$ moments. This provides the unique possibility to study the low temperature electronic in-plane anisotropy of iron pnictides without applying any symmetry-breaking external force.

\begin{figure}[h]
\centering
\includegraphics[width=0.4\textwidth]{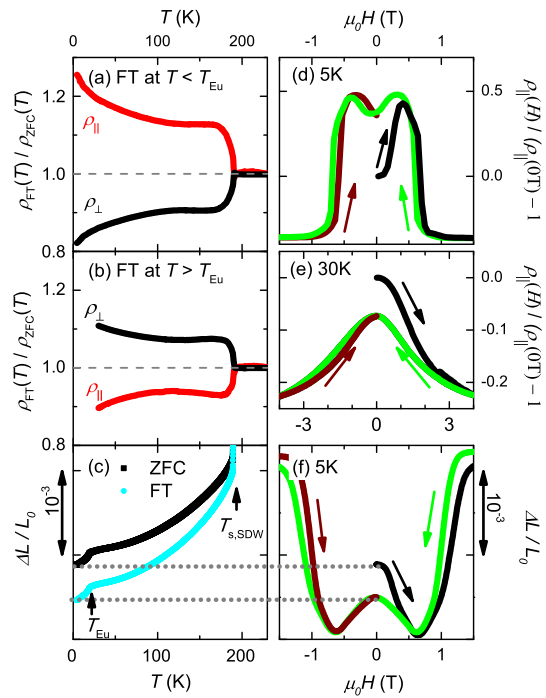}
\caption{(color) (Magneto-)resistance and thermal expansion~/~magnetostriction of ZFC EuFe$_2$As$_2$ ($H \parallel [110]_{\text{T}}$). (a,b)~Temperature-dependent resistivity $\rho_\text{FT}(T)$ after field-treatment~(FT) with 4\,T for currents parallel (red) and perpendicular (black) to $H$, normalized to ZFC $\rho_\text{ZFC}(0\text{T})$. For FT at (a)~$T = 4$\,K and (b)~at 30\,K the opposite behavior is observed. (c)~Thermal expansion $\Delta L (T)/ L_0$ ($\Delta L \parallel H$) after ZFC (black) and FT with 2\,T at 4\,K (cyan). (d,e)~Magnetoresistance $\rho_\parallel(H)/\rho_\parallel(0\,\text{T}) - 1$ at (d)~5\,K and (e)~30\,K as well as (f)~magnetostriction $\Delta L (H)/ L_0$ at 5\,K for increasing (black, brown) and decreasing (green) $H$. All measurements are consistent with a persistent detwinning induced by low magnetic fields that is achievable below or above $T_\text{Eu}$ and robust against heating up to $T_\text{s,SDW}$.}
\label{Fig1} 
\end{figure}

\textit{Resistivity and thermal expansion}
Our remarkable observation is a persistent detwinning after the magnetic field is removed, effective even at temperatures above $T_\text{Eu}$. This can be directly seen in Fig.~\ref{Fig1}a,b, which show the temperature-dependent resistivity $\rho(T)$ after FT, normalized to its ZFC value. Although the magnetic field is switched off before measuring, we observe a strong in-plane anisotropy along the orthorhombic axes. The anisotropy is opposite for FT below and above $T_\text{Eu}$. Furthermore, its magnitude agrees with values obtained for mechanically detwinned Eu compounds~\cite{Ying2011, Jiang13}, and remains virtually constant up to $T_\text{s,SDW}$.

Concurrent evidence for a persistent structural detwinning is also found in thermal expansion, $\Delta L(T) / L$, which is shown in Fig.~\ref{Fig1}c for ZFC and FT. Apart from transitions at  $T_\text{Eu}$ and $T_\text{s,SDW}$, the sample is shorter after FT at $T = 4$\,K than in the ZFC state. Relative length changes are of the order of $10^{-3}$, exceeding the typical magnetostriction of Eu-based materials by at least two orders of magnitude. The induced imbalance of twin domains stays constant up to $T_\text{s,SDW}$ (based on a comparison with experimentally determined lattice constants~\cite{Tegel08}, see Supplementary Information~\cite{Supplementary}).

\textit{Magnetoresistance and magnetostriction}
In order to understand the detwinning mechanism in more detail, we have investigated the field-dependent magnetostriction and magneto-resistance for a ZFC crystal. As the magnetostriction for $T < T_\text{Eu}$ (see Figure~\ref{Fig1}f) directly shows, with increasing field the sample first contracts, and then expands along $H$. After decreasing the field to 0\,T, the original length is not recovered and the sample remains shorter. Assuming $\rho_b > \rho_a$, which was found for mechanically detwinned iron pnictides~\cite{Fisher11}, the same behavior is visible in the magnetoresistance. Thus, the changes in magnetoresistance can be attributed to the magnetic detwinning of the system and not predominantly to electron-spin scattering as suggested by Ref.~\onlinecite{Xiao12}. We conclude from the field-dependent measurements that there are two separate detwinning processes when $T < T_\text{Eu}$. The first occurs at lower fields where the majority of domains get preferentially oriented with the $b$-axis parallel to $H$; the hereby induced imbalance of twin domains persists even when the magnetic field is removed. Secondly, the crystal gets detwinned with $a \parallel H$ at slightly higher fields ($< 1\,$T).

On the other hand, at $T > T_\text{Eu}$ (see Figure~\ref{Fig1}e), with increasing field, the sample only expands along $H$. After decreasing the field to 0\,T, the original length and resistance is again not recovered and the sample remains longer. Thus we conclude that above $T_\text{Eu}$, only the latter detwinning process with $a \parallel H$ takes place.

\begin{figure}[h]
\centering
\includegraphics[width=0.4\textwidth]{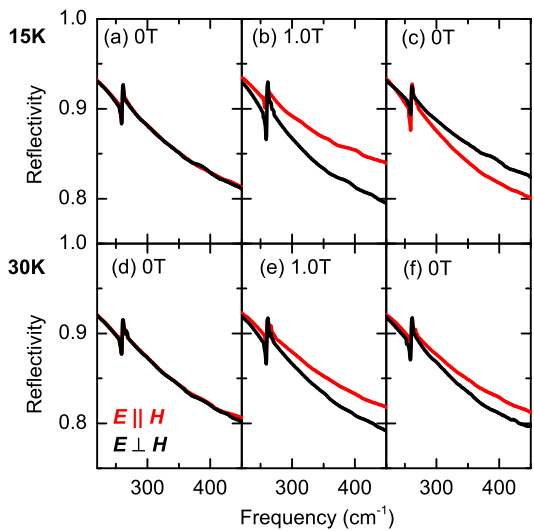}
\caption{(color) Frequency dependent reflectivity of EuFe$_2$As$_2$ at $T=$15 and 30\,K, for $H$ = 0\,T (ZFC), 1\,T and 0\,T after FT ($H \parallel [110]_{\text{T}}$). The magnetic field induces anisotropy between $R(E \parallel H)$ (red) and $R(E \perp H)$ (black) as well as changes the Fe-As phonon mode at $\sim 260\,$cm$^{-1}$, both corresponding to the magnetic detwinning.}
\label{Fig2} 
\end{figure}

\textit{Magneto-optics}
Previous infrared spectroscopy on EuFe$_2$As$_2$ has revealed that Eu$^{2+}$ spin scattering does not notably influence the reflectivity in the far-infrared (FIR) energy range~\cite{Wu09}. Therefore, we have performed low-frequency magneto-optical reflection measurements.
Figure~\ref{Fig2} exhibits polarization-dependent spectra between 220-450\,cm$^{-1}$. We chose $T=15$\,K, 30\,K, as well as $H=0$\,T (ZFC), 1.0\,T, and 0\,T after FT (with 1\,T) as representative. Those frequencies are dominated by the SDW gap and an Fe-As phonon mode at $\sim 260\,\mbox{cm}^{-1}$ (for further information about the gap analysis and mode assignment see Ref.~\onlinecite{Wu09}). 
For 1\,T and after FT, a notable difference is induced between the two polarizations, consistent with the above described magnetic detwinning. 
The origin of the anisotropy is a stronger gap opening and an enhanced phonon oscillator strength along the $b$-axis, which was also found in mechanically detwinned Ba-122 compounds~\cite{Degiorgi11, Nakajima11, Schafgans11,Degiorgi14}.

\begin{figure}[h]
\centering
\includegraphics[width=0.48\textwidth]{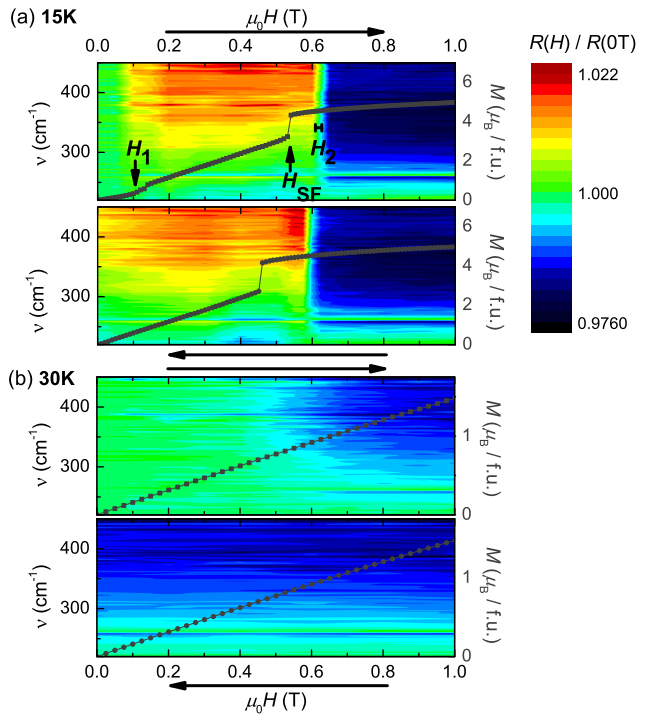}
\caption{(color) EuFe$_2$As$_2$.  Field-dependent magnetization $M(H)$ (grey dots) and false colour plot of the frequency and field-dependent FIR relative reflectivity $R(H)/R(0\,\text{T})$ ($E \perp H$) at (a)~$T=$15\,K and (b)~30\,K ($H \parallel [110]_{\text{T}}$). The detwinning fields $H_1$ and $H_2$ as well as the spin flip field $H_\text{SF}$ of Eu$^{2+}$ can be identified ($H_1 < H_\text{SF} < H_2$).}
\label{Fig3} 
\end{figure}

\begin{figure*} [!http]
\centering
\includegraphics[width=0.7\textwidth]{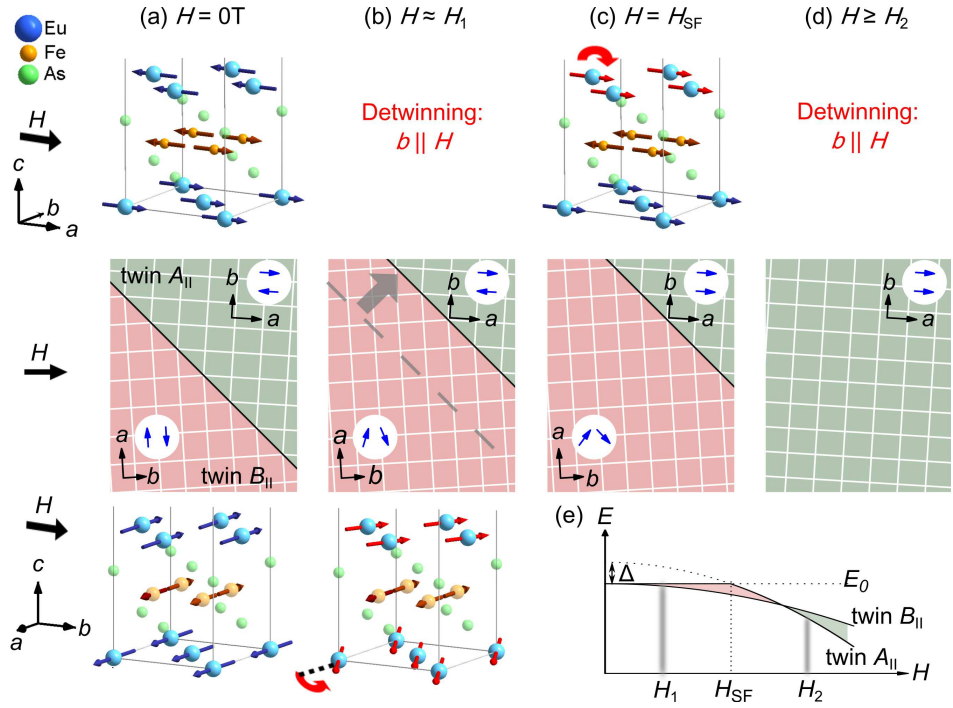}
\caption{(color) Sketch of twin distribution and spin configuration of EuFe$_2$As$_2$  dependent on an (from left to right increasing) external magnetic field $H\parallel [110]_{\text{T}}$ at $T < T_\text{Eu}$. Top and bottom rows show a detail of the EuFe$_2$As$_2$ crystal and magnetic structure (Eu atoms and spins (blue), Fe atoms (yellow) and spins (brown), As atoms (green)) that correspond to the twin distribution sketched in the middle row.
(a)~For $H=0$\,T, the ZFC crystal is twinned and 
the domains are equally distributed. The
Eu$^{2+}$ spins are ordered A-type antiferromagnetically with the spin direction along the $a$-axis.
(b)~With external field, twin~variant~$B_{\parallel}$~(red, bottom) with $b \parallel H$ gets energetically favored and therefore grows on the expense of variant~$A_{\parallel}$~(green, top), as soon as the energy difference exceeds the twin boundary pinning energy.
(c)~With further increasing field, Eu$^{2+}$ spins in the remaining type~$A_{\parallel}$ twins flip along the field direction. Energetically, twin~$B_{\parallel}$ is still more favorable.
(d)~At slightly higher fields, twin~$A_{\parallel}$ is favored and the crystal is detwinned with $a \parallel H$.
(e)~Corresponding energy curves ($E_0$: ground state). While at low fields twin~$B_{\parallel}$ is energetically favorable (red area), twin~$A_{\parallel}$ gets favorable at higher fields (green area). Detwinning takes place at $H_1$ and $H_2$, when the energy gain exceeds the pinning energy of the twin boundary.}
\label{Fig4} 
\end{figure*}

Further information on the twin dynamics can be obtained from the false-colour plot of the field and frequency-dependent relative reflectivity ($E \perp H$) in Fig.~\ref{Fig3}.
For $T = 15$\,K $< T_\text{Eu}$ and increasing magnetic field, $R(H)$ increases rapidly between 0.075 and 0.15\,T, afterwards staying almost constant.
Thus, $H_{1} \sim 0.1\,$T can be identified as the critical field where twins preferentially align with $b\parallel H$.
A sharp drop in the reflectivity at $H_{2} \sim 0.6$\,T marks the second, opposite detwinning process with $a\parallel H$.
For decreasing $H$, the latter process is reversible with a slightly lower critical field. However, the detwinning at low fields is persistent.
At $T = 30$\,K $> T_\text{Eu}$, $R(H)$ continuously decreases with increasing magnetic field, until it saturates at $\sim 0.9$\,T.
With decreasing $H$, the reflectivity stays almost constant and the detwinning with $b \parallel H$ is persistent, even when the field is switched off.

\textit{Magnetization}
Since the Eu$^{2+}$ magnetic moments drive the detwinning (comparison measurements on BaFe$_2$As$_2$ can be found in the Supplementary Information~\cite{Supplementary}), we also include the field-dependent magnetization, which is dominated by Eu$^{2+}$ moments~\cite{Jiang09}, in Fig.~\ref{Fig3}.
At $T = 15$\,K, $M(H)$ exhibits two transitions for increasing, but only the upper one for decreasing field.
The first transition at $\sim 0.1$\,T corresponds to $H_1$ found in reflectivity. However, the second transition, which can be identified due to an abrupt jump in $M(H)$ as a spin flip, precedes the $H_2$ transition observed in reflectivity by about 0.1\,T. 
At $T = 30$\,K, no transition is visible in $M(H)$. Temperature dependent magnetization measurements are consistent with our interpretations and are shown in the Supplementary Information~\cite{Supplementary}.

\textit{Model}
We suggest a simple model to explain the detwinning process above and below $T_\text{Eu}$,
based on the competition between magneto-crystalline anisotropy~$\Delta$, antiferromagnetic exchange coupling~$J$ and Zeeman energy. 
Two twin domains have to be considered, one with the easy $a$-axis perpendicular (type $B_{\parallel}$) and one with it parallel (type $A_{\parallel}$) to $H$ (see Fig.~\ref{Fig4}). After cooling in zero magnetic field, the crystal is twinned with equally distributed variants. At $T < T_\text{Eu}$ and with $H \parallel [110]_{\text{T}}$, minimizing the energy yields for the two twin variants (see Ref.~\onlinecite{Blundell}, Supplementary Information~\cite{Supplementary},  and Fig.~\ref{Fig4}e):
\begin{align}
E_\text{min}^{B_{\parallel}} &= E_0 - \frac{M^2 (\mu_0 H)^2}{2 J M^2 + \Delta}\\
E_\text{min}^{A_{\parallel}} &= E_0 + \Delta - \frac{M^2 (\mu_0 H)^2}{2 J M^2 - \Delta}
\end{align} 
with ground state energy $E_0$. Thus, at low fields, the Eu$^{2+}$ spins of variant~$B_{\parallel}$ gradually rotate towards $H$, lowering the system's energy. When $E_\text{min}^{ B_{\parallel}}-E_\text{min}^{A_{\parallel}}$ exceeds the twin boundary pinning energy, variant~$B_{\parallel}$ grows irreversibly on the expense of variant~$A_{\parallel}$, and the crystal gets detwinned with $b \parallel H$. 
Increasing the magnetic field further induces a spin flip in variant~$A_{\parallel}$, but this twin variant is energetically favorable only at slightly higher fields, when the crystal gets detwinned with $a \parallel H$.
At $T > T_\text{Eu}$, only a detwinning with $a \parallel H$ occurs, because the unordered Eu$^{2+}$ spins align gradually in the direction of the magnetic field. 
As characteristic for domain dynamics, these processes are strongly irreversible, leading to a significant persistent detwinning.

\textit{Concluding discussion}
The question remains why the magnetic detwinning fields are reduced in EuFe$_2$As$_2$ by more than two orders of magnitude (compared to other iron pnictides). Magneto-elastic coupling usually arises due to spin-orbit interactions. However, the orbital momentum of Eu$^{2+}$  is zero; thus the magnetic anisotropy induced by spin-orbit interactions is negligible. Another possibility to induce magnetic anisotropy is by dipole-dipole interactions. However, the resulting anisotropy is much weaker~\cite{Tosti03}. Therefore, other unconventional interactions must cause our observations. From the phase diagram of doped or (chemically) pressurized Eu compounds, it is well known that the Eu$^{2+}$  and Fe$^{2+}$  magnetic orders are strongly intertwined~\cite{Nowik11a, Nowik11b, Zapf11, Akbari11}.  Furthermore, the magnetic moment of Fe$^{2+}$ is non-zero, leading in BaFe$_2$As$_2$ to significant magneto-elastic coupling ~\cite{Chu10}.  Hence, we suggest that the Eu$^{2+}$ spins couple indirectly to the lattice via the Fe$^{2+}$ spins.

We have shown by using resistivity, thermal expansion, magnetostriction, magneto-resistance, magneto-optical and magnetization measurements, that EuFe$_2$As$_2$ can be \emph{persistently} detwinned by laboratory-scale magnetic fields, yielding similar detwinning fractions as commonly used mechanical devices. The detwinning is possible below and above the local Eu$^{2+}$ magnetic ordering, however the mechanism is slightly different: while at $T < T_\text{Eu}$ the crystal gets detwinned with $b \parallel H$ at low fields ($\sim 0.1$\,T at 15\,K) and with $a \parallel H$ at high magnetic fields ($\sim 0.6$\,T at 15\,K), at $T > T_\text{Eu}$, only the latter takes place. We propose that the Eu$^{2+}$ moments couple indirectly to the lattice via the Fe$^{2+}$ spins. To our knowledge, this is the first time that such an indirect coupling has been demonstrated. Most strikingly, a significant imbalance of twin domains remains even when the field is switched off, and the temperature is raised up to $T > T_\text{s,SDW}$.  Such indirect magneto-elastic coupling and its persistence up to much higher energy scales could be interesting also for other materials and even for technical applications.  In summary, the whole effect uncovers a remarkable interdependence between magnetic, electronic and structural effects and allows examining macroscopically the intrinsic in-plane anisotropy of iron pnictides without the application of any symmetry-breaking external force.

We thank N.~Bari\v{s}i\'c, L.~Degiorgi, I.~Mazin, J.~Schmalian and Y.~Xiao for fruitful discussions. The project was supported by DFG SPP 1458 and the German Academic Exchange Service (DAAD).


\begin{thebibliography}{99}
\bibitem{tetra}
In this article, we adapt the notation of the tetragonal system $[hkl]_{\text{T}}$ which is rotated by $45\,^{\circ}$ along the $c$-axis with respect to the orthorhombic system $[hkl]_{\text{O}}$, e.g. $[110]_{\text{T}} = [100]_{\text{O}}$.
\end{thebibliography}
\end{document}